\documentclass[a4paper,fleqn,final]{cas-dc}

\usepackage[T1]{fontenc}
\usepackage[utf8]{inputenc}
\usepackage[main=english]{babel}
\DeclareUnicodeCharacter{03C1}{\ensuremath{\rho}}
\DeclareUnicodeCharacter{03B4}{\ensuremath{\delta}}
\DeclareUnicodeCharacter{03B1}{\ensuremath{\alpha}}
\usepackage{textcomp}
\usepackage{hyperref}
\usepackage{amsmath,amssymb}
\usepackage{graphicx}
\usepackage{booktabs}
\usepackage{tabularx}
\usepackage{array}
\usepackage{calc}

\usepackage{threeparttable}
\usepackage{longtable}
\usepackage{adjustbox}
\usepackage[numbers,sort&compress]{natbib}
\usepackage{caption}
\usepackage{etoolbox}
\usepackage{placeins}
\usepackage{stfloats}
\usepackage{xurl}
\usepackage{microtype}
\usepackage{lastpage}

\makeatletter
\g@addto@macro{\UrlBreaks}{\do\_\do\-\do\/\do\.\do\0\do\1\do\2\do\3\do\4\do\5\do\6\do\7\do\8\do\9}
\makeatother

\renewcommand{\ttfamily}{\rmfamily}
\AtBeginDocument{%
  \urlstyle{same}%
}

\usepackage{xcolor}
\definecolor{linknavy}{RGB}{0,70,127}
\hypersetup{
  colorlinks=true,
  linkcolor=black,
  citecolor=linknavy,
  urlcolor=linknavy,
  runcolor=linknavy
}

\AtBeginEnvironment{table}{\let\sffamily\rmfamily}
\AtBeginEnvironment{figure}{\let\sffamily\rmfamily}
\AtBeginEnvironment{table*}{\let\sffamily\rmfamily}
\AtBeginEnvironment{figure*}{\let\sffamily\rmfamily}

\ExplSyntaxOn
\RenewDocumentCommand \firstname {}
  { \textcolor{black}{\seq_use:Nn \l_stm_au_seq { ~ }} }

\RenewDocumentCommand \emailauthor { m m }
   {
     \int_gincr:N \g_ead_int
     \seq_gput_right:Nn \g_stm_ead_seq
       {
         { \href{mailto:#1}{\rmfamily #1} }
         \parsename { #2 }
         \space(\eadauthor)
       }
     }

\cs_set:Npn \__first_footerline:
{
  \group_begin:
  \small
  \normalfont
  \ifnum\theblind>0\relax
  \else
  \__short_authors: :~
  \fi
  \itshape Preprint~submitted~to~Elsevier
  \group_end:
}

\RenewDocumentCommand \printorcid { } { }

\cs_set:Npn \__first_head:
{
  \parbox[t]{\textwidth}
  {
    \rule{\textwidth}{0pt}
  }
}

\cs_set:Npn \__cas_head:
{
  \parbox{\textwidth}
  {
    \rule{\textwidth}{0pt}
  }
}

\cs_set:Npn \__cas_foot:
{
  \parbox[t]{\textwidth}
  {
   \rule{\textwidth}{.2pt}\\
   \small
   \normalfont
   \__first_footerline:
   \hfill Page~\thepage {}~of~ \lastpage
  }
}
\ExplSyntaxOff

\makeatletter
\ps@cas
\makeatother

\begin{document}
% ============================================================

% IEEE BST control: show full author names (no dashes for repeated names)
\makeatletter
\def\bstctlcite{\@ifnextchar[{\@bstctlcite}{\@bstctlcite[@auxout]}}
\def\@bstctlcite[#1]#2{\@bsphack
  \@for\@citeb:=#2\do{%
    \edef\@citeb{\expandafter\@firstofone\@citeb}%
    \if@filesw\immediate\write\csname #1\endcsname{\string\citation{\@citeb}}\fi}%
  \@esphack}
\makeatother
\bstctlcite{IEEEbsTcontrol}

\shortauthors{Mitra et al.}
\shorttitle{Sentinel-2 active-fire dataset}

\title[mode=title]{A Sentinel-2 benchmark dataset for deep-learning active-fire
  segmentation across 25 California wildfires, July 2020--August 2026}

\author[1]{Shreyan Mitra}
\author[2]{Mohammadreza Narimani}
\cormark[1]
\ead{mnarimani@ucdavis.edu}
\author[2]{Parastoo Farajpoor}

\affiliation[1]{organization={California High School},
  city={San Ramon},state={CA},postcode={94583},country={USA}}
\affiliation[2]{organization={Department of Biological and Agricultural
  Engineering, University of California, Davis},
  city={Davis},state={CA},postcode={95616},country={USA}}

\cortext[1]{Corresponding author}

% ------------------------------------------------------------------
\begin{abstract}
This article describes an open image dataset for developing and evaluating
active-fire segmentation methods in satellite imagery.  The dataset contains
2{,}148 image-mask pairs from 25 California wildfires, with acquisitions
spanning July~2020 to August~2026.  Each image is a $512\times512$-pixel,
three-channel composite derived from Sentinel-2 Level-2A bands B12, B11 and
B8A at 20\,m spatial sampling.  A fixed linear rendering is applied throughout
the dataset.  Corresponding masks distinguish background, SWIR-rule active fire
and invalid observations.  The masks were generated from shortwave-infrared
brightness and near-infrared contrast, followed by constrained neighborhood
growth.  The release includes chip-level metadata and an incident-disjoint
partition containing 18 training, three validation and four test fires.  Among
the image pairs, 841 contain active-fire labels; these labels occupy
0.0766\% of all grid cells.  A mask-blind analyst review covers 233 test chips
and provides a separate assessment of the rule-generated labels at chip and
connected-component levels.  Reference training and evaluation code accompanies
the data, including a ResNet-34 U-Net implementation with validation-based
checkpoint and threshold selection.  The archived images, masks, metadata and
review annotations support research on rare-class segmentation, learning from
algorithmic labels and transfer across fire incidents.  The versioned dataset
is deposited on Zenodo, with preparation and reuse software maintained in a
public GitHub repository.
\end{abstract}


\maketitle

% ------------------------------------------------------------------
% Specifications Table (unnumbered; Data in Brief convention)
% ------------------------------------------------------------------
\begin{table*}[t]
\caption{Specifications table.}
\label{tab:specs}
\centering
\begin{tabularx}{\textwidth}{@{} >{\raggedright\arraybackslash}p{0.22\textwidth} X @{}}
\toprule
\textbf{Item} & \textbf{Description} \\
\midrule
Subject
  & Earth and Planetary Sciences \\[2pt]
Specific subject area
  & Satellite remote sensing; active-fire segmentation; deep learning with
    algorithmic supervision \\[2pt]
Type of data
  & Three-channel images; categorical masks; tabular metadata; analyst
    annotations; descriptive figures; Python software \\[2pt]
How the data were acquired
  & Sentinel-2 Level-2A assets were accessed through the Earth Search
    SpatioTemporal Asset Catalog (STAC) service.  National Interagency Fire
    Center (NIFC) Wildland Fire Interagency Geospatial Services (WFIGS)
    perimeter records supplied incident locations and dates.  Images were
    sampled, rendered and labeled with the public preparation code. \\[2pt]
Data format
  & Processed 8-bit RGB PNG images; single-channel integer PNG masks; CSV
    metadata.  Source reflectance assets are accessed during construction;
    they are not the distributed image representation. \\[2pt]
Description of data collection
  & Twenty-five large California wildfire incidents
    (WFIGS \texttt{attr\_IncidentTypeCategory\,=\,WF}; all retained names
    verified as WF) were selected from the queried records.  Up to 14
    acquisition dates per incident were sampled within bounded post-discovery
    windows.  Scene-specific $1{,}536\times1{,}536$-pixel windows were divided
    into $512\times512$-pixel chips.  The final archive contains both
    fire-bearing and fire-free chips. \\[2pt]
Data source location
  & California, United States.  The incident distribution is shown in
    Figure~\ref{fig:locations}.  Dataset preparation and analysis: University
    of California, Davis, Davis, CA 95616, USA\@.  Acquisition dates reported
    for the release: 23~July~2020 to 21~August~2026. \\[2pt]
Data accessibility
  & Repository: Zenodo.  Dataset: California Sentinel-2 Active-Fire
    Segmentation Dataset, version 1.0.0 \citep{mitra2026dataset}.
    DOI: \href{https://doi.org/10.5281/zenodo.22713948}{10.5281/zenodo.22713948}.
    Data license: CC~BY~4.0. \\[2pt]
Companion software
  & \texttt{California\_Sentinel2\_Active\_Fire\_Dataset}
    \citep{mitra2026code}, MIT license.  Public repository:
    \url{https://github.com/MohammadrezaNarimaniUCDavis/California_Sentinel2_Active_Fire_Dataset}.
    The implementation described here was inspected at commit
    \texttt{e5e4667}. \\[2pt]
Related research article
  & None. \\
\bottomrule
\end{tabularx}
\end{table*}

% ==================================================================
\section{Value of the Data}\label{sec:value}
% ==================================================================

\begin{itemize}
\item The paired images and masks provide a common input representation for
  active-fire segmentation, including fire-free scenes and an explicitly encoded
  invalid-observation class.

\item Remote-sensing researchers, computer-vision researchers and wildfire
  scientists can use the incident-disjoint partition to compare methods on fires
  excluded from training.

\item Sparse positive labels and a wide range of hotspot sizes support studies
  of class imbalance, sampling, decision thresholds and learning from
  algorithmic supervision.

\item The separately distributed analyst annotations support assessment of label
  agreement and development of review or relabeling strategies without replacing
  the original rule-generated targets.

\item Open preparation and training code documents the rendering, label rule,
  sampling and evaluation conventions, allowing users to distinguish
  reproduction of the archived experiment from construction of an updated
  dataset.
\end{itemize}

% ==================================================================
\section{Background}\label{sec:background}
% ==================================================================

An active-fire mask describes where a satellite observation contains evidence
of ongoing combustion; it is not a map of the area eventually burned.
Sentinel-2 provides shortwave-infrared observations at 20\,m sampling that can
be used to resolve spatial patterns of active fire \citep{drusch2012sentinel2}.
Converting those observations into reusable learning data requires an explicit
record of how images were selected, how labels were generated and which
observations belong to each evaluation partition.

Related Sentinel-2 research in crop-yield estimation highlights how feature
representation, reference-data availability and transfer across sites and years
influence model use \citep{narimani2026cropyield}.  An active-fire benchmark
addresses a different target, but benefits from the same explicit separation of
sensor observations, derived labels and evaluation units.

Existing resources address related tasks at different observation scales.
Pereira et al.\ \citep{pereira2021activefire} developed a Landsat-8 active-fire
dataset with algorithmic annotations and a separate manually annotated
evaluation resource.  Fire-video annotations provide frame-level segmentation
data for camera-based surveillance \citep{wahyono2022fireimages}, while
geospatial incident datasets describe the timing, location and extent of
wildfire and prescribed-fire activity \citep{beidler2024geospatial}.  The
present dataset complements these resources with fixed-rendering Sentinel-2
image-mask pairs, incident-level partitions and a separately documented analyst
review.  Its purpose is to provide an accessible reference dataset for pixel
classification under sparse, algorithmically generated supervision, rather than
to introduce a new segmentation architecture.

% ==================================================================
\section{Data Description}\label{sec:data}
% ==================================================================

% ------------------------------------------------------------------
\subsection{Release organization and partition inventory}\label{sec:partition}
% ------------------------------------------------------------------

\begin{figure*}[t]
\centering
\includegraphics[width=\textwidth]{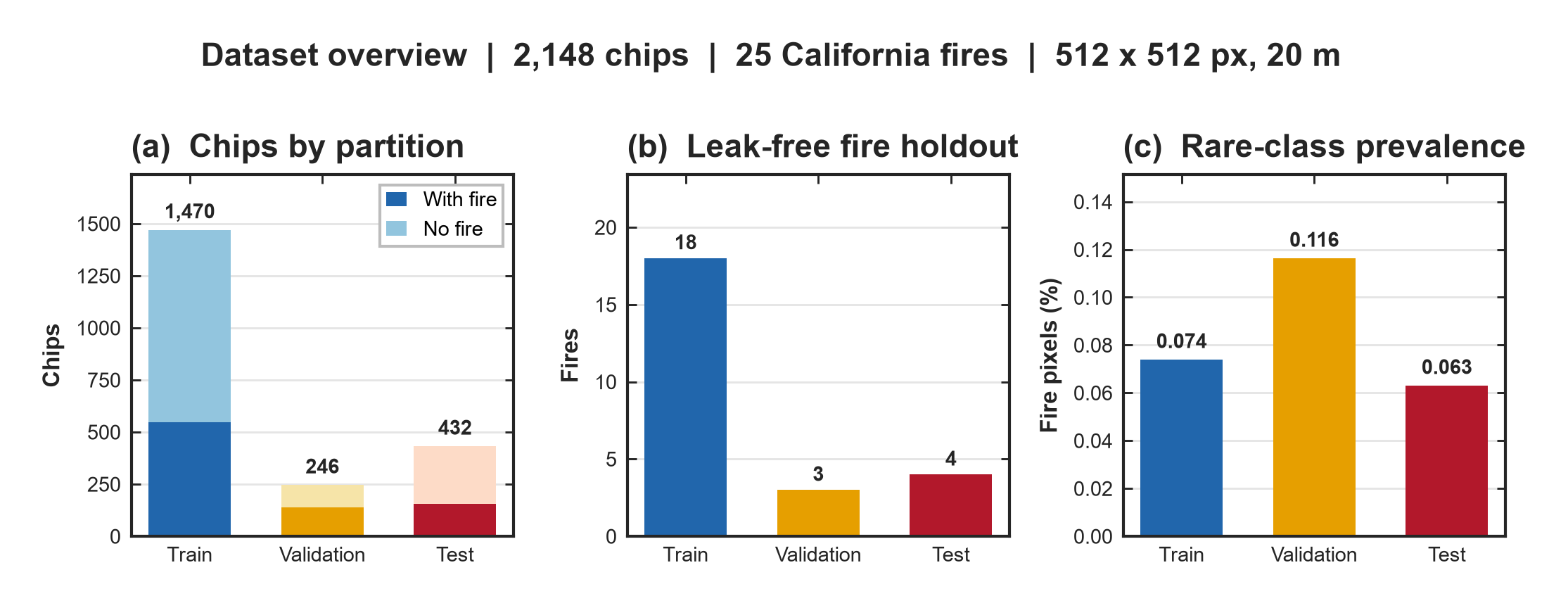}
\caption{Dataset and partition inventory.  (a)~Chip counts, subdivided into
  chips with and without rule-generated fire labels.  (b)~Numbers of incidents
  in the training, validation and test partitions.  (c)~Fire-pixel prevalence
  as a percentage of all grid cells in each partition.  Dark and light portions
  in panel~(a) denote fire-bearing and fire-free chips, respectively.  The
  partition is disjoint by incident name.}
\label{fig:overview}
\end{figure*}

The dataset is archived as version 1.0.0 on Zenodo \citep{mitra2026dataset}.
The deposit comprises paired image and mask directories, a chip manifest, a
partition file and analyst-review masks.  The public repository
\citep{mitra2026code} provides the construction, training and
annotation-import scripts.  Table~\ref{tab:partition} gives the partition
inventory; Figure~\ref{fig:overview} summarizes the corresponding chip and
incident counts and the proportion of grid cells labeled as fire.

\begin{table*}[t]
\caption{Incident-disjoint partition inventory.  Fire prevalence uses all
  $512\times512$ grid cells in the denominator, including cells encoded as
  invalid; training and evaluation instead exclude invalid cells.}
\label{tab:partition}
\centering
\begin{tabular*}{\textwidth}{@{\extracolsep{\fill}} l r r r r r @{}}
\toprule
Partition & Incidents & Chips & Fire-bearing chips & Fire pixels &
  Fire prevalence (\%) \\
\midrule
Training   & 18 & 1{,}470 & 548 & 284{,}948 & 0.0739 \\
Validation &  3 &   246   & 137 &  75{,}028 & 0.1163 \\
Test       &  4 &   432   & 156 &  71{,}451 & 0.0631 \\
\midrule
Total      & 25 & 2{,}148 & 841 & 431{,}427 & 0.0766 \\
\bottomrule
\end{tabular*}
\end{table*}

All observations assigned to Castle, McCash and Bobcat belong to validation.
Slater, Hopkins, CALDWELL and Windy form the test partition; the remaining 18
incidents form the training partition.  The archive contains 1{,}307 fire-free
chips and 841 fire-bearing chips, so the released corpus is not a balanced
binary image-classification dataset.  These counts describe the archived
samples, not an estimate of the frequency of fire in California satellite
observations.

Each image and its corresponding mask share a basename of the form
\texttt{\{fire\_slug\}\_\{YYYY-MM-DD\}\_r\{row\}c\{column\}.png}.  The row and
column indices range from 0 to~2 within a sampled parent window.  A file can
therefore be associated with its incident, observation date and position in
that window without opening the image.

% ------------------------------------------------------------------
\subsection{Spatial and temporal coverage}\label{sec:spatial}
% ------------------------------------------------------------------

\begin{figure*}[t]
\centering
\includegraphics[width=\textwidth,height=0.62\textheight,keepaspectratio]{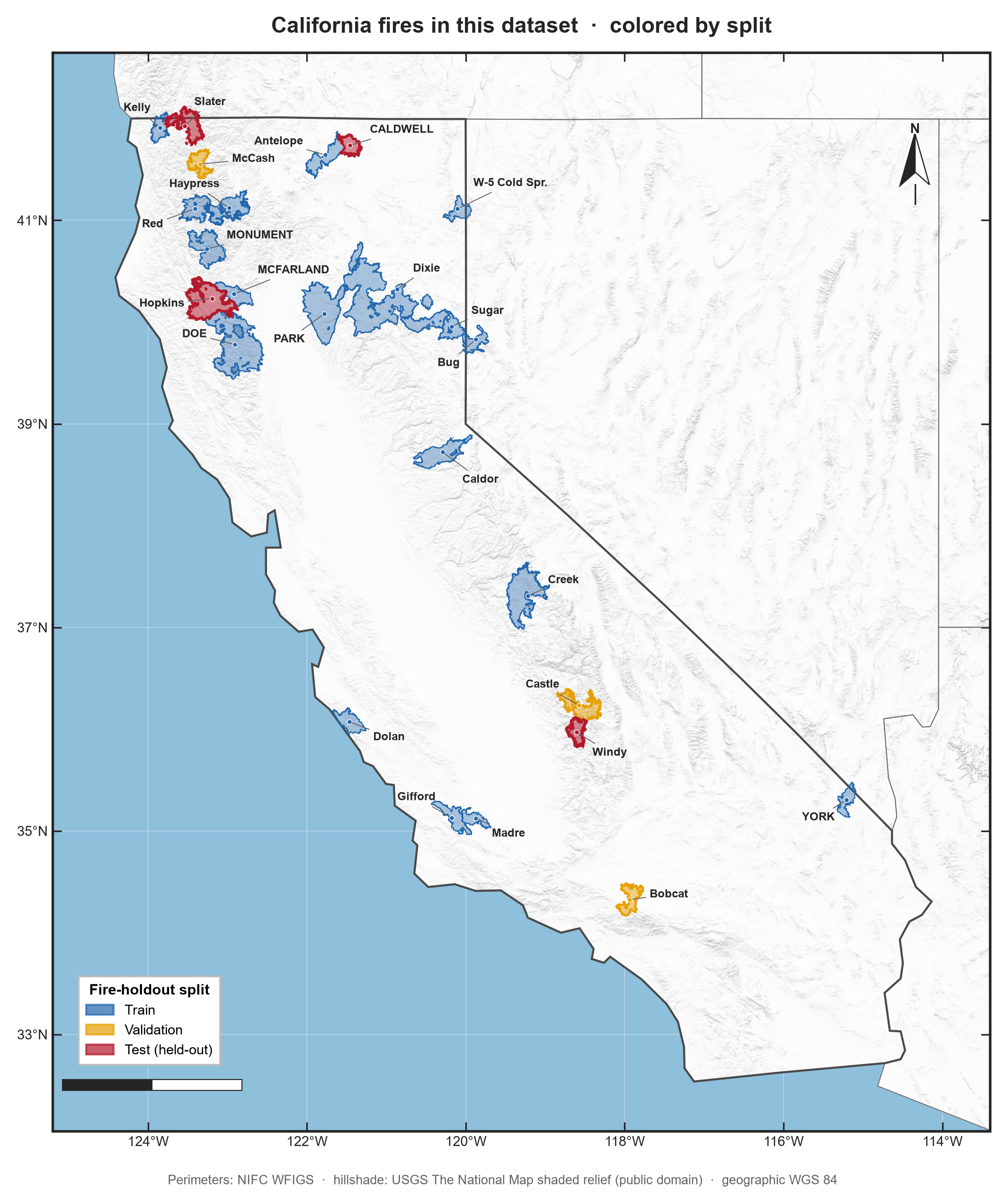}
\caption{Locations of the California incidents, colored by partition: training,
  validation and test.  Perimeters are from NIFC WFIGS; the background is USGS
  The National Map shaded relief (public domain).  Geographic coordinates
  provide spatial reference.  Perimeters depict incident extents, not
  instantaneous flame fronts.  Map lines delineate study areas and do not
  necessarily depict accepted national boundaries.}
\label{fig:locations}
\end{figure*}

Figure~\ref{fig:locations} locates the 25 incident perimeters and their
partition assignments.  Perimeters provide the spatial context for sampling;
they are not the active-fire labels.  Individual chips sample a 10.24\,km-wide
window, while each sampled parent window spans 30.72\,km.  The release includes
repeated observations of incident surroundings, rather than a continuous
statewide mosaic.

Repeated acquisitions can support different analytical tasks.  For example,
Sentinel-2 time-series analysis of broomrape-infested tomato fields used
growing-degree-day alignment to compare vegetation trajectories across crop
stages \citep{narimani2025broomrape}.  Here, dates identify individual fire
observations within bounded incident windows; the distributed chips support
acquisition-specific segmentation rather than a phenologically aligned sequence.

% ------------------------------------------------------------------
\subsection{Image channels, label encoding and metadata}\label{sec:channels}
% ------------------------------------------------------------------

The distributed RGB channels represent B12, B11 and B8A, respectively.  The
first two are shortwave-infrared bands and the third is the narrow
near-infrared band.  These three bands are sampled at their native 20\,m
spacing \citep{drusch2012sentinel2}.  A single set of rendering limits is
applied to all chips, rather than estimating a separate contrast stretch for
each image.  Table~\ref{tab:files} summarizes the files and metadata needed to
read the release.

At a different observation scale, multi-trait grapevine modeling used
hyperspectral measurements to estimate leaf biochemical and nutritional
attributes \citep{farajpoor2025grapevine}.  The distinction is one of
representation and target: quantitative trait retrieval uses spectral
measurements, whereas these rendered image channels provide a common input for
locating rule-labeled fire.

\begin{figure*}[t]
\centering
\includegraphics[width=\textwidth]{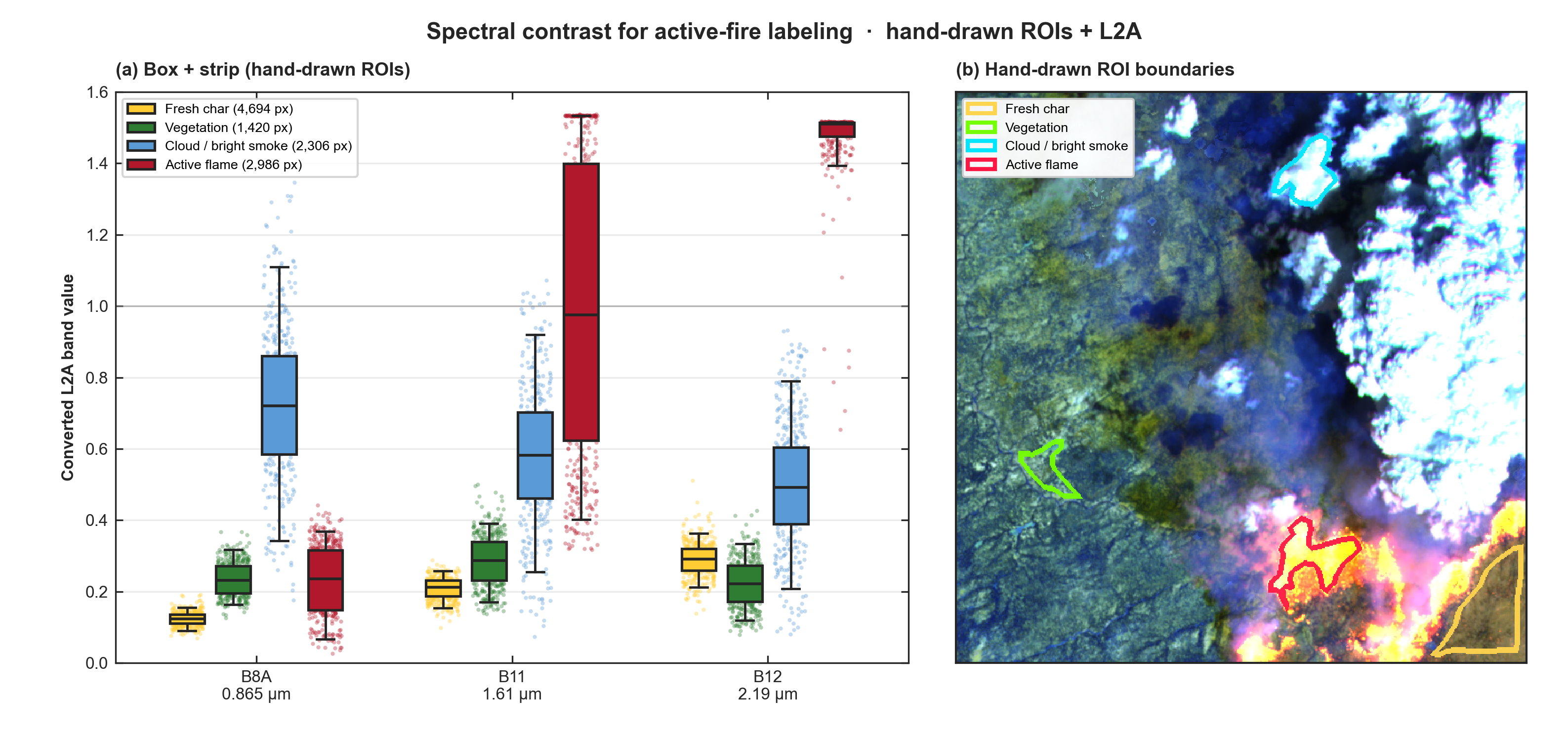}
\caption{Spectral contrast associated with active-fire labeling.
  (a)~Distributions of converted Sentinel-2 Level-2A B8A, B11, and B12 band
  values extracted from hand-drawn regions representing fresh char, vegetation,
  cloud/bright smoke, and active flame in a Castle-fire image.
  (b)~Locations of the corresponding hand-drawn regions on the SWIR composite.
  Values represent discrete band measurements rather than continuous spectral
  signatures.  The algorithmic fire-label rule is defined separately in
  Eqs.~(\ref{eq:contrast})--(\ref{eq:growth}).}
\label{fig:spectral}
\end{figure*}

\begin{figure*}[t]
\centering
\includegraphics[width=\textwidth,height=0.72\textheight,keepaspectratio]{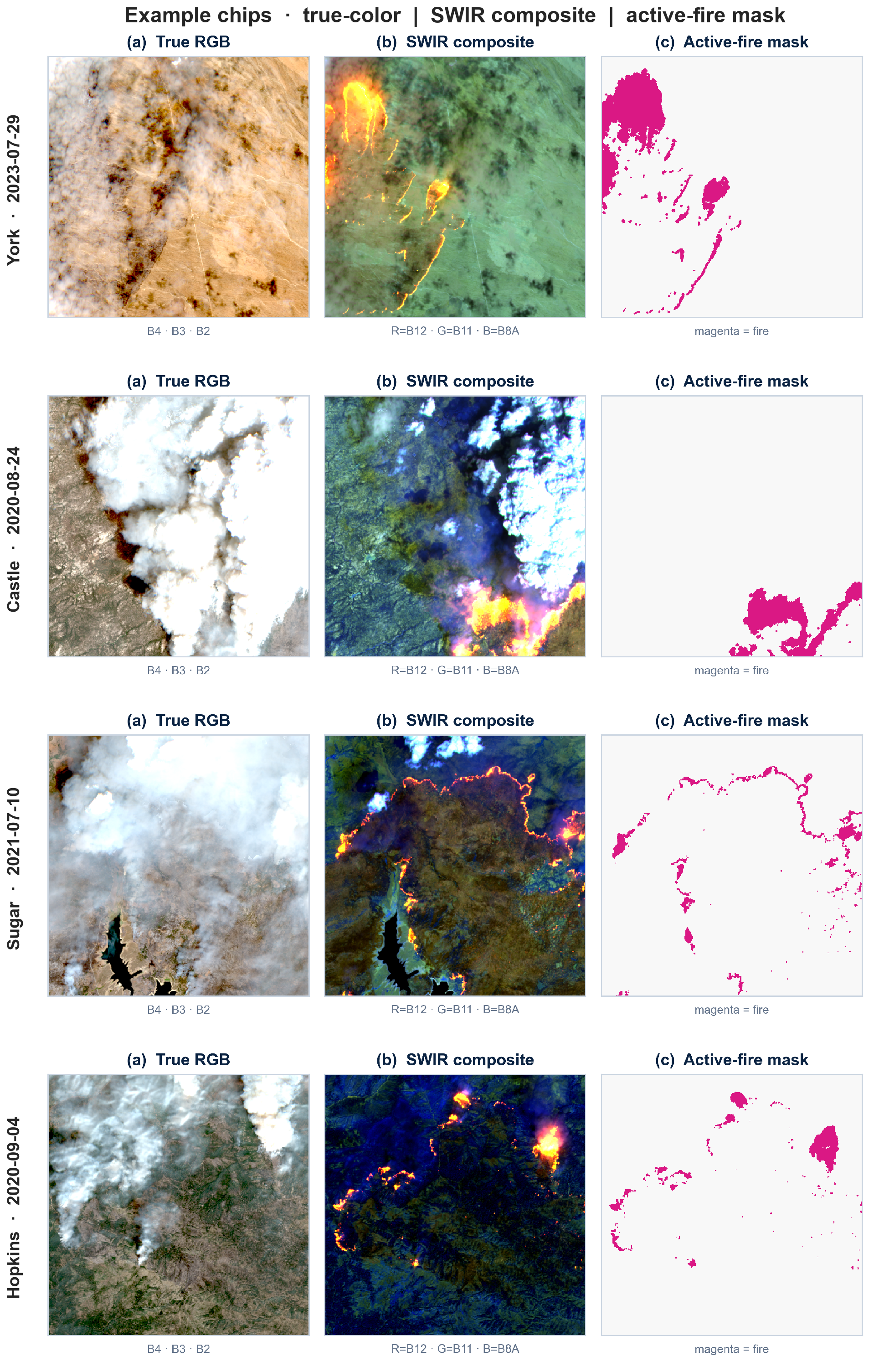}
\caption{Examples from York (29~July~2023), Castle (24~August~2020), Sugar
  (10~July~2021) and Hopkins (4~September~2020).  Columns show true color
  (B4-B3-B2), the SWIR composite (B12-B11-B8A) and the rule-generated fire
  mask in magenta.  True-color panels are contextual illustrations, not
  additional channels in the released three-channel PNGs.  Display colors do
  not change the integer encoding of the distributed masks.}
\label{fig:examples}
\end{figure*}

The spectral rationale for the label rule follows the use of shortwave-infrared
contrast for hot-target detection \citep{murphy2016hotmap}.
Figure~\ref{fig:spectral} illustrates that contrast with measured converted L2A
band-value distributions for the three distributed bands (B8A, B11, B12) on
hand-drawn class ROIs from a Castle chip.  Figure~\ref{fig:examples} presents
four image-mask examples, with true-color views included solely as visual
context.

\begin{table*}[t]
\caption{File and metadata dictionary.  The \texttt{summary.csv} table has one
  row per distributed chip.  Partition membership is supplied separately and is
  also derived from incident names by the training script.}
\label{tab:files}
\centering
\begin{tabularx}{\textwidth}{@{} p{0.30\textwidth} X @{}}
\toprule
\textbf{File or field} & \textbf{Interpretation} \\
\midrule
\texttt{images/*.png}
  & $512\times512\times3$ arrays, unsigned 8-bit integers; channels B12,
    B11, B8A. \\[2pt]
\texttt{masks/*.png}
  & $512\times512$ arrays; 0\,=\,background, 1\,=\,SWIR-rule active fire,
    255\,=\,invalid observation. \\[2pt]
\texttt{summary.csv}: \texttt{file\_name}
  & Shared image-mask basename. \\[2pt]
\texttt{summary.csv}: \texttt{fire}; \texttt{date}
  & Incident name; acquisition date in YYYY-MM-DD format. \\[2pt]
\texttt{summary.csv}: \texttt{day\_of\_burn}
  & Integer days since the recorded discovery date; not a direct measure of
    local combustion duration. \\[2pt]
\texttt{summary.csv}: \texttt{fire\_px}; \texttt{nodata\_pct}
  & Number of value-1 mask cells; percentage of cells encoded 255. \\[2pt]
\texttt{partitions.csv}
  & Assignment of incident names to training, validation or test. \\[2pt]
\texttt{analyst\_review/hand\_masks/}
  & Review masks for 233 test chips; 255 also represents analyst
    uncertainty. \\
\bottomrule
\end{tabularx}
\end{table*}

% ------------------------------------------------------------------
\subsection{Positive-class prevalence and spatial structure}\label{sec:prevalence}
% ------------------------------------------------------------------

\begin{figure*}[t]
\centering
\includegraphics[width=\textwidth]{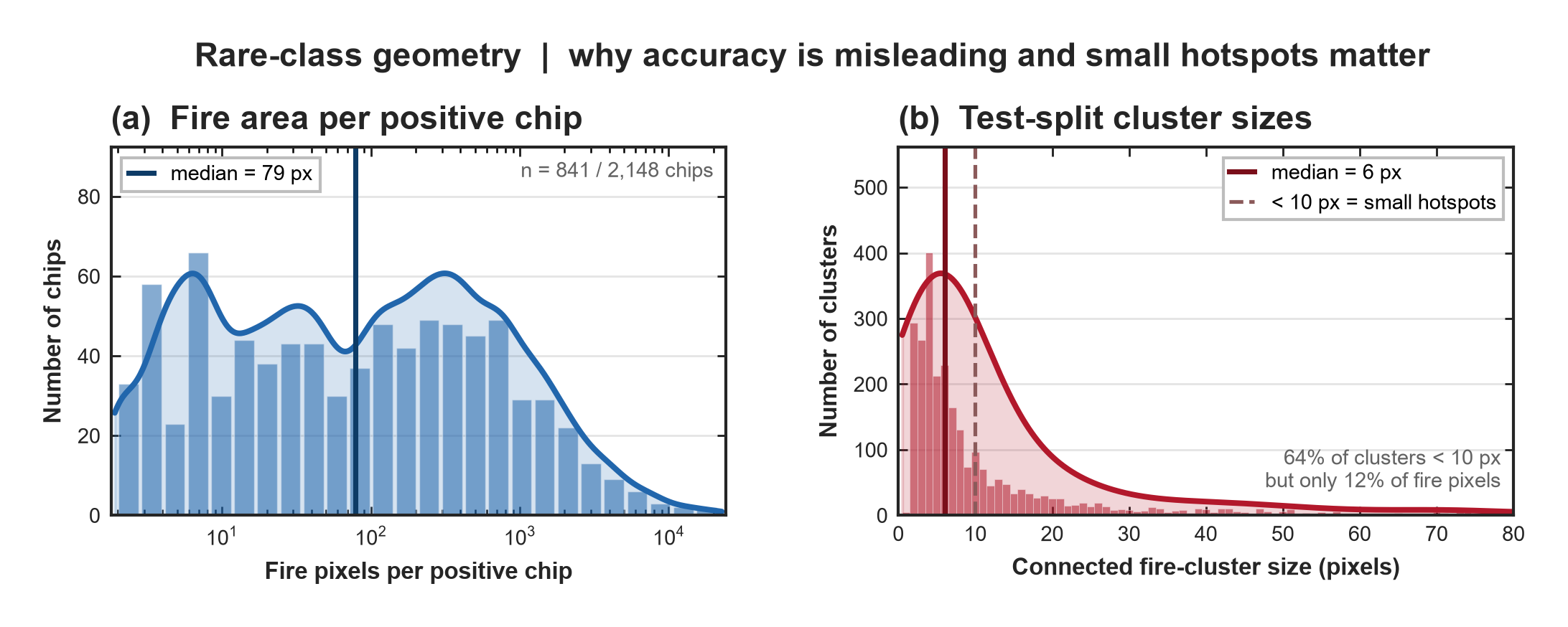}
\caption{Spatial structure of the positive class.  (a)~Distribution of
  fire-pixel counts across the 841 fire-bearing chips; the horizontal axis is
  logarithmic and the vertical marker denotes the median of 79 pixels.
  (b)~Distribution of connected fire-component sizes in the test partition,
  displayed up to 80 pixels.  The solid and dashed markers indicate the
  six-pixel median and the ten-pixel size reference.  The inset percentages
  refer to all test components and fire-labeled pixels, not only the displayed
  size range.}
\label{fig:imbalance}
\end{figure*}

Across the release, 431{,}427 grid cells carry active-fire labels.  A total of
841 chips, or 39.2\% of all chips, contain at least one such cell.  The median
fire-bearing chip contains 79 fire-labeled cells.
Figure~\ref{fig:imbalance}a describes the distribution across positive chips,
and Figure~\ref{fig:imbalance}b describes connected-component sizes in the test
partition.

The median test component contains six pixels.  Components smaller than ten
pixels account for approximately 64\% of test components but 12\% of
test fire-labeled pixels.  Component counts and pixel counts thus emphasize
different aspects of the same targets.  The inventory supports both pixel-level
evaluation and separately defined component-level evaluation, provided that the
connectivity and matching conventions are stated.

Related optical canopy mapping has reported candidate-center agreement
separately from pixel precision and overlap, because confirmation of a detected
object does not establish agreement across its full boundary
\citep{narimani2026canopy}.  In this release, component confirmation and
pixel-wise model agreement are likewise reported as different quantities rather
than interchangeable measures of accuracy.

At 20\,m sampling, one grid cell represents a nominal 400\,m$^{2}$ ground
support.  That support must not be interpreted as the physical area occupied by
flame within the cell.  Counts summed across chips and acquisition dates are
pixel-observations: the same ground location can contribute more than once.
They are therefore neither unique burned-area totals nor estimates of combustion
area.

% ------------------------------------------------------------------
\subsection{Reference training and label-review outputs}\label{sec:outputs}
% ------------------------------------------------------------------

\begin{figure*}[t]
\centering
\includegraphics[width=\textwidth]{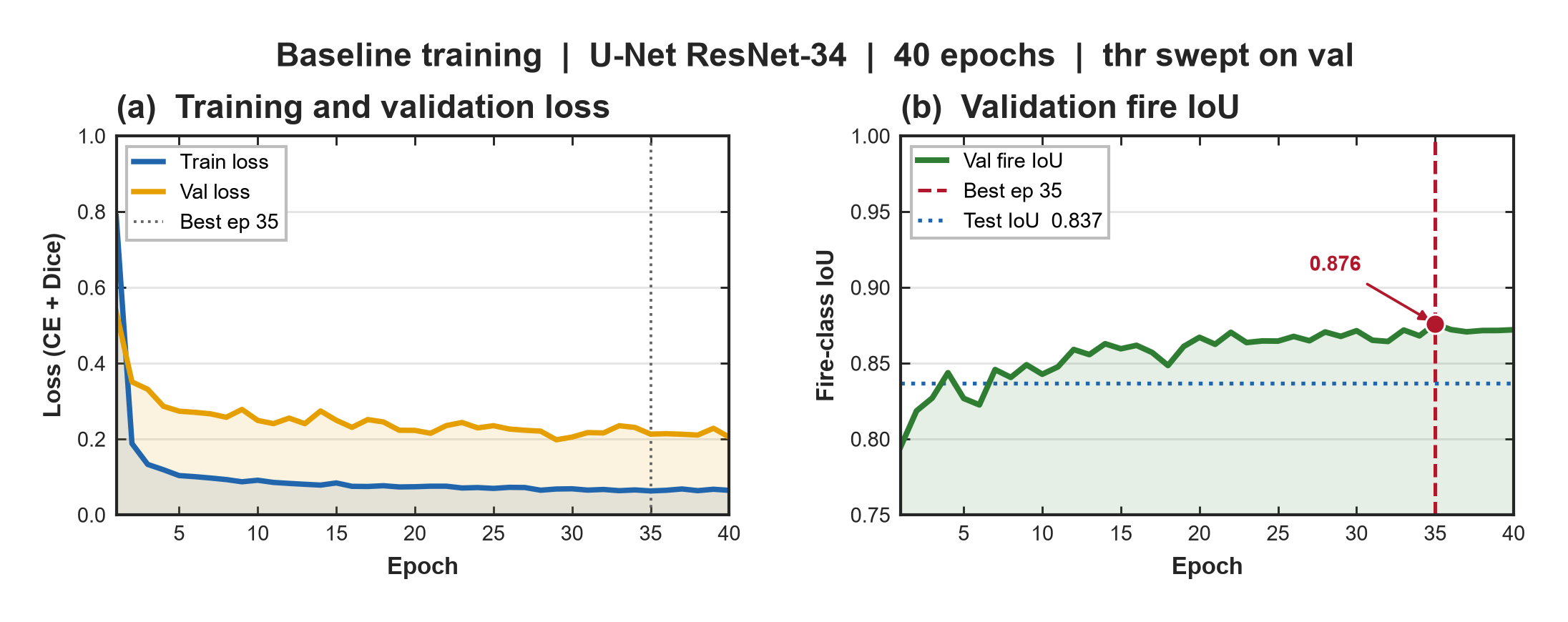}
\caption{Reported 40-epoch reference run.  (a)~Training and validation losses
  for the weighted cross-entropy plus soft-Dice objective.  (b)~Validation
  fire-class intersection over union (IoU), using the threshold selected on
  validation at each epoch.  The selected checkpoint is epoch~35, with
  validation IoU 0.876.  The horizontal test reference, IoU 0.837, is shown
  for context and is not a checkpoint-selection criterion.}
\label{fig:training}
\end{figure*}

\begin{figure*}[t]
\centering
\includegraphics[width=\textwidth]{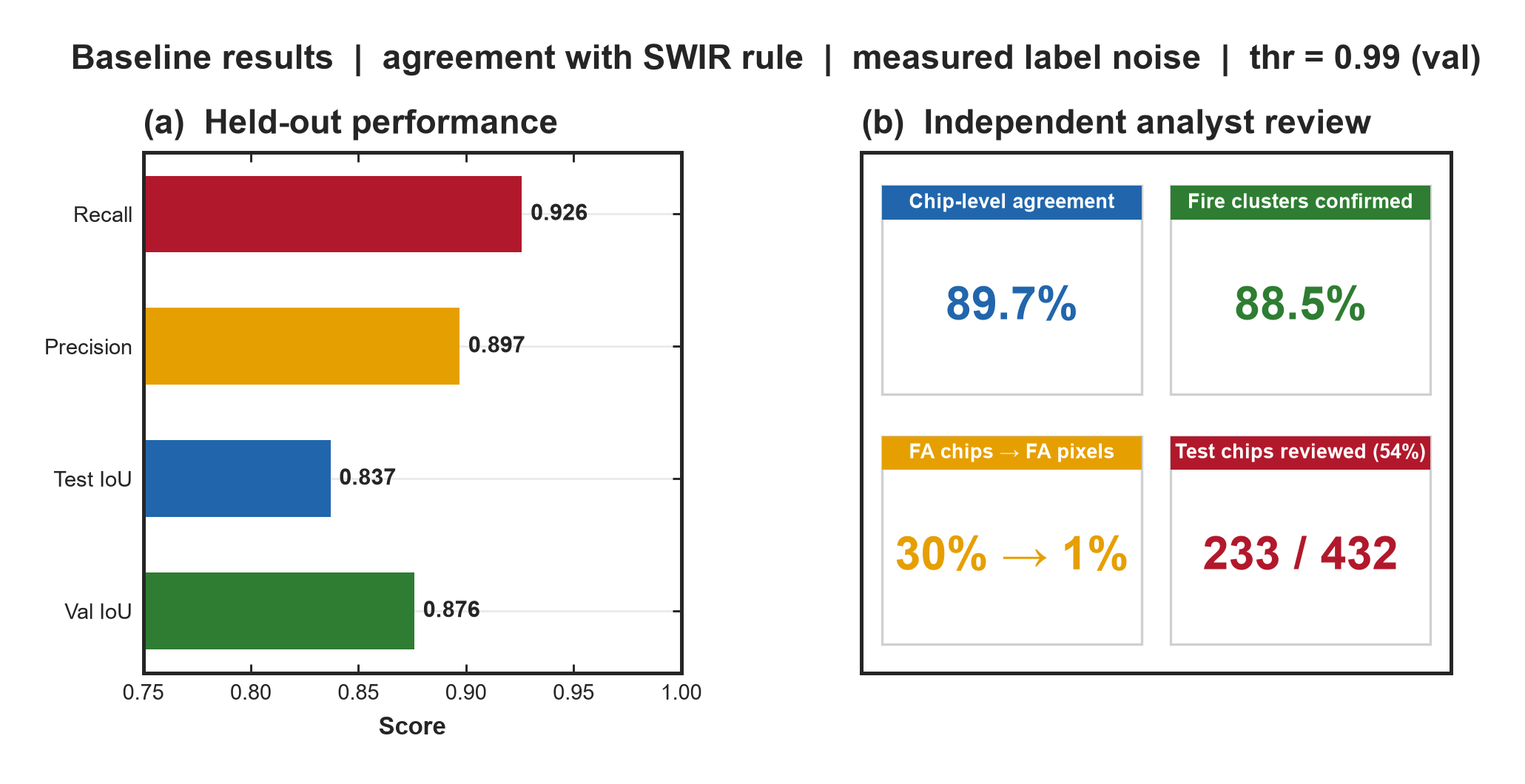}
\caption{Reference-model agreement and analyst-review summary.  (a)~Validation
  IoU and test IoU, precision and recall against the SWIR-rule masks; the score
  axis begins at 0.75.  (b)~Review of 233 of 432 test chips.  Chip agreement
  is 89.7\%; 88.5\% of rule components overlap an analyst fire annotation.
  Approximately 30\% of reviewed rule-positive chips were visually
  unconfirmed; those chips contain approximately 1\% of the rule-positive
  pixels in the reviewed batch.  These two percentages have different
  denominators and do not describe a before-after reduction.}
\label{fig:performance}
\end{figure*}

\begin{figure*}[t]
\centering
\includegraphics[width=\textwidth]{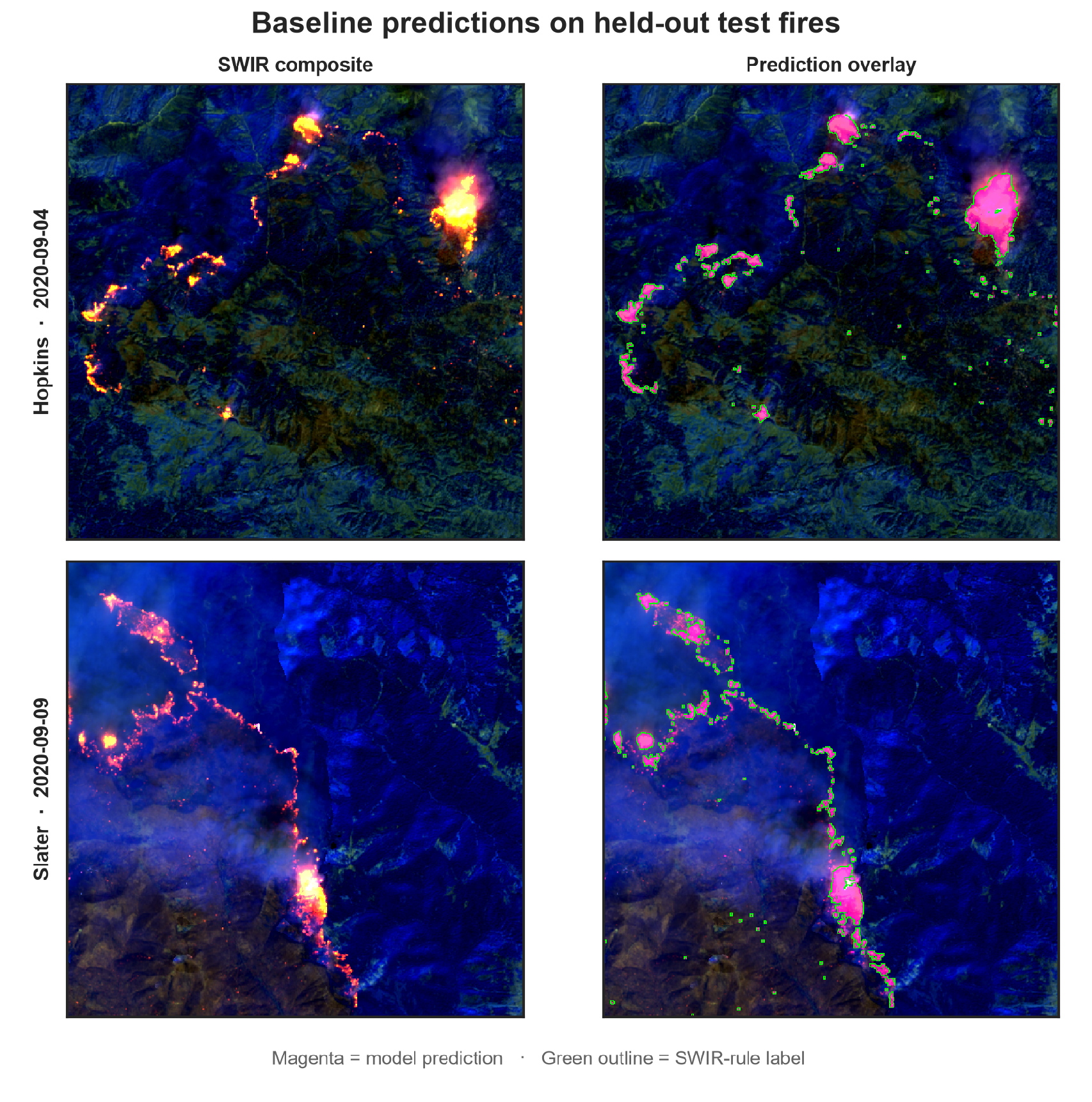}
\caption{Reference predictions for Hopkins (4~September~2020) and Slater
  (9~September~2020).  Left: SWIR composite.  Right: the same view with model
  predictions shown in magenta and the SWIR-rule boundary in green.  The green
  outlines identify the algorithmic reference, not an independent field or
  analyst delineation.  Colors are visualization overlays; training uses the
  original composites and integer masks.}
\label{fig:predictions}
\end{figure*}

The companion implementation supplies a reference route from the PNG pairs to
segmentation predictions \citep{mitra2026code}.
Figures~\ref{fig:training}--\ref{fig:predictions} document the reported
reference run and the analyst-review summary.  These outputs demonstrate use of
the dataset under a specified partition and evaluation procedure; they are not a
comparison with other architectures or independent operational fire-detection
measurements.

At the selected checkpoint, the validation-derived probability threshold is
0.99.  Applying that operating point to the four test incidents gives a pooled
fire-class IoU of 0.837, precision of 0.897 and recall of 0.926 against the
distributed SWIR-rule masks (Figure~\ref{fig:performance}a).  The score refers
to the positive class and excludes invalid observations.

The analyst review is separate from the reference-model evaluation.  It covers
233 test chips (53.9\%) and reports agreement with the algorithmic labels at
chip and component levels (Figure~\ref{fig:performance}b).  Its sampling,
uncertainty handling and component-confirmation criterion are described in
Section~\ref{sec:review}.  Figure~\ref{fig:predictions} juxtaposes the model
predictions and rule masks for two held-out incidents.

% ------------------------------------------------------------------
\subsection{Incident contributions}\label{sec:contributions}
% ------------------------------------------------------------------

\begin{figure*}[t]
\centering
\includegraphics[width=\textwidth]{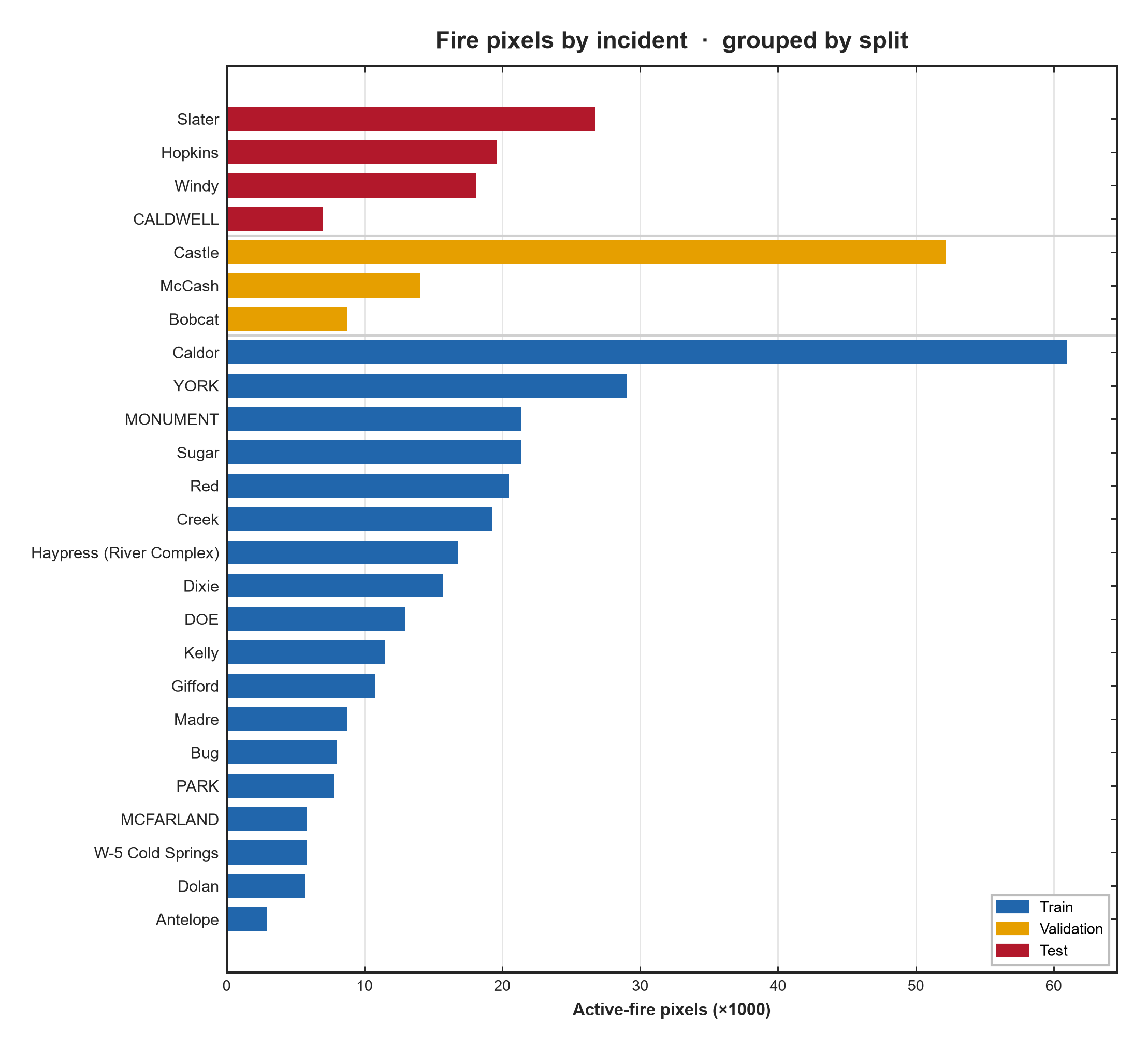}
\caption{Contributions of active-fire pixel-observations by incident, grouped
  into test, validation and training partitions and ordered within each group.
  The axis is expressed in thousands of labeled pixels.  Counts are accumulated
  over the sampled chips and dates; they do not represent unique burned area or
  a ranking of total incident size.}
\label{fig:contributions}
\end{figure*}

Figure~\ref{fig:contributions} provides the incident-level inventory of
active-fire pixel-observations, grouped by partition.  Contributions vary with
incident extent, sampled dates, cloud or smoke conditions and the location of
active fronts relative to the sampling window.  Reporting this distribution
allows reuse studies to distinguish pooled pixel agreement from evaluations that
give equal weight to each incident.

% ==================================================================
\section{Experimental Design, Materials and Methods}\label{sec:methods}
% ==================================================================

% ------------------------------------------------------------------
\subsection{Incident selection and acquisition sampling}\label{sec:incidents}
% ------------------------------------------------------------------

Incident records were obtained from the WFIGS Interagency Perimeters service
\citep{nifc2026wfigs}.  The query selected California wildfire records
(\texttt{attr\_IncidentTypeCategory\,=\,WF}; prescribed-fire RX excluded)
with discovery dates after 1~January~2019 and polygon areas greater than
20{,}000 acres, equivalent to approximately 80.94\,km$^{2}$.  Records were
ordered by polygon area; repeated incident names were reduced to the largest
returned polygon, and 25 incidents were retained.  All 25 retained names were
verified as WF on the WFIGS service.  This procedure defines the queried sample
and should not be interpreted as a complete census of all California fires.

For each incident, the sampling point was the midpoint of its geographic
bounding box.  The observation window began at the recorded discovery date and
ended at containment or 60 days after discovery, whichever occurred first.
When containment was unavailable, the code used 45 days after discovery; a
seven-day minimum was enforced.  Table~\ref{tab:acquisition} lists the
acquisition and preparation settings used by the public implementation.

Sentinel-2 Level-2A scenes were queried through Earth Search
\citep{element84earthsearch}.  The search intersected the incident sampling
point and allowed scene-level cloud cover up to 95\%, thereby retaining
acquisitions in which parts of the incident might remain visible.  For
duplicate dates, the scene with the lowest reported cloud cover was retained.
Up to 14 dates were selected at approximately even index intervals from the
chronological list of available scenes, rather than at fixed calendar
intervals.  The build procedure does not apply the scene-classification cloud
mask to remove cloud and smoke pixels from the labels or images.

\begin{table*}[t]
\caption{Acquisition and preparation settings in the public implementation.}
\label{tab:acquisition}
\centering
\begin{tabularx}{\textwidth}{@{} p{0.22\textwidth} X @{}}
\toprule
\textbf{Step} & \textbf{Setting or convention} \\
\midrule
Incident source
  & WFIGS Interagency Perimeters; California wildfire records only
    (\texttt{attr\_IncidentTypeCategory\,=\,WF}; prescribed-fire RX
    excluded); post-1 January 2019 discovery; area ${>}80.94$\,km$^{2}$;
    25 retained names. \\[2pt]
Date window
  & Discovery to containment, capped at 60 days; 45-day fallback and
    seven-day minimum. \\[2pt]
Scene source
  & Earth Search \texttt{sentinel-2-l2a}; point-intersection search; scene
    cloud cover ${\leq}95$\%. \\[2pt]
Scene selection
  & Lowest-cloud scene per date; at most 14 dates per incident. \\[2pt]
Sampling grid
  & $1{,}536\times1{,}536$ cells at 20\,m, centered on the incident
    bounding-box midpoint and snapped in the selected scene's projected
    coordinate system. \\[2pt]
Chip geometry
  & Non-overlapping $3\times3$ subdivision into $512\times512$-cell chips
    within each parent window. \\[2pt]
Footprint screen
  & Discard a chip if more than 50\% of its cells are invalid. \\[2pt]
Negative-chip selection
  & Retain all positive chips; apply integer-stride subsampling to the
    negative-chip list.  Final archive: 841 positive and 1{,}307 negative
    chips. \\
\bottomrule
\end{tabularx}
\end{table*}

% ------------------------------------------------------------------
\subsection{Radiometric conversion and image rendering}\label{sec:rendering}
% ------------------------------------------------------------------

Bands B8A, B11 and B12 were read on the selected scene's 20\,m grid.  Each
band used the scale and offset supplied in its STAC asset metadata, with the
implementation's fallback values of 0.0001 and zero when those fields were
absent.  Digital number zero was treated as a source no-data sentinel and reset
to zero after conversion.  The image and mask for a given chip share the same
grid.  Across dates, scene-specific projected coordinate systems are retained;
the PNG archive is not a temporally coregistered multiband geospatial stack.

The radiometric conversion and fixed rendering can be written as
\begin{equation}\label{eq:rendering}
\rho_b = s_b D_b + o_b,\qquad
I_b = \mathrm{round}\!\left[255\,\mathrm{clip}\!\left(\frac{\rho_b}{u_b},
      0, 1\right)\right]
\end{equation}
In Eq.~(\ref{eq:rendering}), $D_b$ is the source digital number in band~$b$,
$s_b$ and $o_b$ are that asset's scale and offset, and $\rho_b$ is the
resulting dimensionless Level-2A value for nonzero source digital numbers.  The
function $\mathrm{clip}(z,0,1)$ restricts $z$ to the interval from zero to
one; $I_b$ is the stored 8-bit channel value.  The upper limits $u_b$ are 0.60
for B12, 0.50 for B11 and 0.45 for B8A.  Channel order is B12-B11-B8A.  These
limits are display transformations, not the thresholds used to generate the
labels.  For a hot target, the converted optical-band signal can include emitted
radiance and should not be interpreted as passive-surface reflectance alone.

% ------------------------------------------------------------------
\subsection{Algorithmic label generation}\label{sec:labels}
% ------------------------------------------------------------------

The detector is a dataset-specific SWIR hot-target rule motivated by HOTMAP
\citep{murphy2016hotmap}, not an unchanged implementation of the published
algorithm.  It operates on the converted band arrays before their clipping and
quantization into PNG channels.  Let the SWIR-to-NIR contrast be
\begin{equation}\label{eq:contrast}
q = \begin{cases}
  \rho_{12}/\rho_{8\mathrm{A}}, & \rho_{8\mathrm{A}} > 10^{-6} \\
  +\infty,                      & \rho_{8\mathrm{A}} \leq 10^{-6}
\end{cases}
\end{equation}
Here $\rho_{12}$ and $\rho_{8\mathrm{A}}$ denote the converted B12 and B8A
values at a pixel.  The small denominator cutoff in Eq.~(\ref{eq:contrast})
specifies the implementation's numerical convention; it is not an independently
validated physical boundary.

The seed set $C$ and the more permissive candidate set $P$ are
\begin{align}
C &= \{\rho_{12} \geq 0.35 \;\text{and}\; q \geq 4.0\}, \label{eq:seeds}\\
P &= \{\rho_{12} \geq 0.25 \;\text{and}\; q \geq 2.0\}. \nonumber
\end{align}
Thus, $C$ contains high-contrast seed pixels and $P$ defines where those seeds
may expand.  All thresholds refer to the converted values, not to the 0--255
rendered channels.  B11 supplies image context and participates in the validity
test, but no B12/B11 condition is used in Eq.~(\ref{eq:seeds}).

Growth is limited to two four-neighbor steps:
\begin{equation}\label{eq:growth}
G_0 = C,\qquad
G_{k+1} = G_k \cup \bigl(\delta_4(G_k) \cap P\bigr),\quad k = 0,1
\end{equation}
In Eq.~(\ref{eq:growth}), $G_k$ is the grown set after $k$ steps and
$\delta_4$ is a one-pixel dilation using horizontal and vertical neighbors.
The intersection with $P$ constrains each step to candidate pixels.  An
eight-connected binary sieve with a two-pixel size parameter is then applied to
$G_2$ using \texttt{rasterio.features.sieve} \citep{mitra2026code}.  This is a
bounded two-step operation, not unconstrained growth through an entire connected
region.

A cell is treated as valid when each of the three converted band values is
nonzero, following the released implementation.  The final mask assigns~1 to
detected fire in valid cells, 0 to remaining valid cells and 255 to cells
failing that validity test.  Accordingly, 255 is a pipeline validity code,
predominantly associated with missing acquisition coverage; it is not a cloud
class.  A valid cell labeled~0 means that the rule did not identify active fire
there, not that independent observations established the absence of combustion.

% ------------------------------------------------------------------
\subsection{Tiling, sampling and evaluation partitions}\label{sec:tiling}
% ------------------------------------------------------------------

Each parent window is subdivided into nine non-overlapping chips.  Chips
exceeding the invalid-cell threshold in Table~\ref{tab:acquisition} are
discarded before packaging.  All fire-bearing chips are retained.  The
negative-chip list is subsampled using an integer stride calculated from the
positive and negative counts \citep{mitra2026code}.  Although the configuration
uses a nominal negative-to-positive ratio of one, integer-stride selection does
not enforce exact equality; the final counts are those in Table~\ref{tab:partition}.

Partition membership is assigned by incident name, using the fixed validation
and test name sets in Section~\ref{sec:partition}.  Image normalization and
training-only augmentation are estimated or drawn from the training partition.
Holding out groups is appropriate when observations within a group share spatial
or temporal structure \citep{roberts2017crossval}.  Here, the grouping prevents
observations carrying the same incident name from appearing in multiple
partitions; it does not by itself certify a geographic buffer between distinct
incidents.

Spatial validation also matters within wildfire applications: analysis of
structure loss in the Palisades Fire compared random and spatial-block
validation to distinguish local prediction from transfer across neighborhoods
\citep{farajpoor2026palisades}.  The incident-level grouping used here targets
transfer across fires; it answers a related, but not identical, generalization
question.

The archived manifest should be used for comparisons on this release.
Re-running queries against evolving source services creates a new acquisition
selection and need not reproduce the same files byte for byte.  For geospatial
applications, users should also retain the source scene identifiers, coordinate
reference systems and grid transforms during reconstruction: these are not
embedded in the distributed PNG representation or its six-field chip manifest.

% ------------------------------------------------------------------
\subsection{Analyst-review protocol and agreement measures}\label{sec:review}
% ------------------------------------------------------------------

The review batch was drawn from the four test incidents and included both
rule-positive and rule-negative chips.  Tasks were presented in a
fire-stratified randomized order.  The analyst inspected SWIR composites
without the generated masks or their fire-pixel counts.  Training-chip examples
were used to explain the annotation convention.  The protocol asked the analyst
to identify visible active-fire cores, leave nonfire regions unmarked and use
an uncertainty label where a reliable judgment could not be made
\citep{mitra2026code}.

The initial labeling canvas used a twofold nearest-neighbor enlargement.
Exported brush masks were mapped back to $512\times512$ cells.  A native cell
was assigned to fire when any corresponding higher-resolution subcell was marked
as fire; uncertainty took precedence where the classes overlapped.  The
importer also used the annotation export to retain reviewed empty chips.  For
comparisons, cells marked 255 by either the rule mask or the analyst mask were
excluded.

Chip agreement compares the presence or absence of at least one fire cell in
the two masks over the retained cells.  A rule component is confirmed when any
of its cells overlaps an analyst fire annotation.  Components use eight-neighbor
connectivity, with no additional spatial tolerance in the matching step
\citep{mitra2026code}.  The reported component-confirmation fraction is
\begin{equation}\label{eq:component}
A_{\mathrm{component}} = \frac{1}{K}\sum_{j=1}^{K}
  \mathbf{1}\bigl[C_j \cap H \neq \emptyset\bigr]
\end{equation}
In Eq.~(\ref{eq:component}), $K$ is the number of rule components across the
reviewed, comparable cells, $C_j$ is the set of cells in component~$j$, $H$ is
the analyst fire-cell set in the corresponding chip, and $\mathbf{1}[\cdot]$ is
an indicator equal to one when the condition is true and zero otherwise.  This
criterion evaluates overlap with an analyst delineation; it does not require
one-to-one correspondence between components and can depend on delineation
width.

The 233-chip review yielded 89.7\% chip agreement and 88.5\% rule-component
confirmation.  The approximately 30\% visually unconfirmed fraction uses
reviewed rule-positive chips as its denominator.  The approximately 1\% pixel
fraction uses all rule-positive pixels in the reviewed batch as its denominator
and counts those lying in visually unconfirmed chips.  It is not an estimate
that only 1\% of all labeled pixels are incorrect.  These descriptive review
results apply to the reviewed subset; they are not field-confirmed detection
probabilities for all test observations.

% ------------------------------------------------------------------
\subsection{Reference segmentation configuration}\label{sec:config}
% ------------------------------------------------------------------

A U-Net \citep{ronneberger2015unet} with a ResNet-34 encoder
\citep{he2016resnet} supplies the reference model.  The implementation supports
ImageNet initialization \citep{deng2009imagenet} and two output classes, with
invalid cells excluded from optimization.  Table~\ref{tab:reference} records
the distributed reference settings.  The model is included as a reuse example
rather than as a claim of architectural novelty.

The encoder-decoder formulation is also used in other dense geospatial labeling
tasks.  A residual U-Net has been applied to farmland-extent mapping from NAIP
imagery, where a separate text-prompted model was explored for boundary
refinement \citep{narimani2026farmland}.  Here, the reference network operates
directly on the three-channel fire composites, providing a documented starting
point for subsequent model comparisons.

\begin{table*}[t]
\caption{Distributed reference training and evaluation settings.  These are the
  public script's configuration and the operating point reported for
  Figures~\ref{fig:training}--\ref{fig:predictions}.  A new training run
  should record its actual backend, software versions and saved checkpoint.}
\label{tab:reference}
\centering
\begin{tabularx}{\textwidth}{@{} p{0.22\textwidth} X @{}}
\toprule
\textbf{Component} & \textbf{Setting} \\
\midrule
Input and output
  & Three rendered channels, normalized with training-derived statistics; two
    class logits. \\[2pt]
Architecture
  & ResNet-34 U-Net implemented with the torchvision fallback used for the
    reported run (\texttt{segmentation\_models\_pytorch} is preferred by the
    script when installed, but was not the backend for the archived
    checkpoint).  ImageNet initialization via torchvision
    \texttt{IMAGENET1K\_V1}. \\[2pt]
Optimization
  & AdamW \citep{loshchilov2019adamw}; initial learning rate 0.0003; weight
    decay 0.0001; cosine learning-rate annealing. \\[2pt]
Training schedule
  & 40 epochs; batch size 8; encoder frozen for two epochs, then unfrozen
    at 0.1 times the initial base learning rate. \\[2pt]
Sampling
  & Replacement sampling; weight~3 for positive chips and weight~1 for
    negative chips; four draws per training chip per epoch on average. \\[2pt]
Geometric augmentation
  & Flips and 90-degree rotations; crop scale 0.55--1.35; fire-centered
    crops with probability 0.85 when a positive cell is available. \\[2pt]
Copy-paste
  & Probability 0.5; one to three donor chips; two-pixel feathered context
    halo; donors restricted to training chips. \\[2pt]
Photometric augmentation
  & Common-channel gain within $\pm12$\%; additive noise with standard
    deviation 0.02 on the $[0,1]$ image scale. \\[2pt]
Loss
  & Class-weighted cross-entropy plus fire-class soft Dice, with Dice
    weight~1 and smoothing constant~1. \\[2pt]
Numerical settings
  & Mixed precision on CUDA; gradient-norm clipping at~5; configured
    seed~0.  Reported run: Kaggle NVIDIA Tesla T4; exact container
    environment documented in \texttt{docs/REFERENCE\_RUN.md}. \\[2pt]
Selection
  & Maximum pooled validation fire IoU over checkpoints and a fixed
    threshold grid. \\[2pt]
Reported operating point
  & Epoch~35; probability threshold 0.99; validation fire IoU 0.876. \\
\bottomrule
\end{tabularx}
\end{table*}

Training samples are augmented on the fly; the distributed images and masks are
not overwritten.  Copy-paste follows the general use of composited labeled
instances for augmentation \citep{ghiasi2021copypaste}, adapted here to fire
regions with a surrounding blending halo.  Donors are drawn only from positive
training chips.  Mask handling preserves categorical values; when zooming out,
the code uses coverage-based retention to avoid dropping sparse positive cells
\citep{mitra2026code}.

The optimization objective combines weighted cross-entropy with a soft-Dice
overlap term, following the use of overlap-based losses for segmentation
\citep{milletari2016vnet}:
\begin{equation}\label{eq:loss}
\mathcal{L} = -\frac{1}{|V|}\sum_{i \in V} w_{y_i} \log p_{i,y_i}
  + 1 - \frac{2\displaystyle\sum_{i \in V} p_{i,1}\,y_i + 1}%
             {\displaystyle\sum_{i \in V} p_{i,1}
              + \sum_{i \in V} y_i + 1}
\end{equation}
Here $V$ is the set of valid cells in the batch; $|V|$ is its size;
$y_i \in \{0,1\}$ is the rule label; and $p_{i,c}$ is the softmax probability
for class~$c$ at cell~$i$.  The cross-entropy weight is $w_0 = 1$ for
background and $w_1 = \min\!\bigl(50,\max\!\bigl(1,\sqrt{1/f_s}\bigr)\bigr)$
for fire, where $f_s$ is the expected fire-cell fraction under the training
sampler before augmentation.  The constants of one in the overlap ratio are the
implementation's smoothing terms.  Equation~(\ref{eq:loss}) specifies the
released linear-sum Dice denominator explicitly.

% ------------------------------------------------------------------
\subsection{Checkpoint selection and reference evaluation}\label{sec:eval}
% ------------------------------------------------------------------

At each epoch, validation probabilities are evaluated at thresholds 0.05,
0.10, \ldots, 0.90, together with 0.95, 0.97, 0.98, 0.99 and 0.995.  The
checkpoint and threshold giving the highest pooled validation fire-class IoU are
retained.  The reported test scores use that validation-selected operating
point, without selecting a threshold to maximize the reported test result.

For all valid cells pooled across the evaluated partition,
\begin{align}
\mathrm{IoU} &= \frac{TP}{TP+FP+FN}, \label{eq:metrics}\\
\mathrm{Precision} &= \frac{TP}{TP+FP}, \nonumber\\
\mathrm{Recall} &= \frac{TP}{TP+FN} \nonumber
\end{align}
In Eq.~(\ref{eq:metrics}), $TP$ counts cells positive in both the model
prediction and the rule mask, $FP$ counts predicted fire in rule-background
cells, and $FN$ counts rule-fire cells predicted as background.  A model cell
is positive when its fire probability is at least the selected threshold.
Counts are accumulated across the partition before division; the reported IoU is
not a mean of per-chip IoUs.  These quantities measure model agreement with the
algorithmic reference and are distinct from the analyst-review quantities in
Eq.~(\ref{eq:component}).

% ==================================================================
\section*{Limitations}
% ==================================================================

The release is intended for image-based learning under documented algorithmic
supervision.  Its active-fire labels are derived from the same optical bands
represented in the inputs; model agreement with those labels is therefore a
measure of reproducing the reference rule.  The mask-blind analyst review adds
a separate visual assessment for 233 test chips but does not establish
field-confirmed pixel boundaries or inter-analyst reliability.

The sampled incidents are large California fires, and the fixed parent windows
need not include their complete perimeters.  Holding out incident names does not
establish separation of every neighboring landscape.  Cloud and smoke can
obscure active fronts, and the optical acquisitions do not sample continuous or
diurnal fire evolution.  The 8-bit rendered channels are suitable for the
documented image-learning task, but not a substitute for the original
radiometric assets in quantitative spectral analysis.  PNGs also omit
geospatial transforms.  Studies requiring geographic transfer, radiometric
retrieval or operational detection should reconstruct the appropriate source
representation and define independent validation suited to that task.

% ==================================================================
\section*{Ethics Statements}
% ==================================================================

This work uses satellite imagery and publicly accessible incident-perimeter
records.  It did not involve human participants, animal experiments or
personally identifiable research data.  Analyst annotations are research
data-curation outputs.  Source imagery and perimeter attribution are retained in
the dataset documentation and figure captions.

% ==================================================================
\section*{CRediT Author Statement}
% ==================================================================

\textbf{Shreyan Mitra:} Conceptualization, Methodology, Software, Validation,
Formal analysis, Investigation, Data curation, Writing -- original draft,
Visualization.
\textbf{Mohammadreza Narimani:} Writing -- review \& editing, Visualization,
Project administration.
\textbf{Parastoo Farajpoor:} Writing -- review \& editing, Supervision.

% ==================================================================
\section*{Acknowledgments}
% ==================================================================

The authors acknowledge the Copernicus Sentinel-2 data program, the Earth
Search service and NIFC WFIGS for access to the source observations and
incident records.  USGS The National Map shaded relief provides the terrain
context in Figure~\ref{fig:locations}.

% ==================================================================
\section*{Funding}
% ==================================================================

This research did not receive any specific grant from funding agencies in the
public, commercial, or not-for-profit sectors.

% ==================================================================
\section*{Declaration of Competing Interest}
% ==================================================================

The authors declare that they have no known competing financial interests or
personal relationships that could have appeared to influence the work reported
in this paper.

% ==================================================================
\section*{Generative AI Statement}
% ==================================================================

During the preparation of this work, the authors used ChatGPT to assist with
language editing, literature contextualization and refinement of visualization
code.  The authors reviewed and edited all resulting content and take full
responsibility for the content of the article.

% ==================================================================
\section*{Data Availability}
% ==================================================================

The image-mask corpus and associated metadata are available as California
Sentinel-2 Active-Fire Segmentation Dataset, version 1.0.0, at
\url{https://doi.org/10.5281/zenodo.22713948} \citep{mitra2026dataset}.  The
companion preparation, training and review software is available at
\url{https://github.com/MohammadrezaNarimaniUCDavis/California_Sentinel2_Active_Fire_Dataset}
\citep{mitra2026code}.  Data are distributed under CC~BY~4.0 and code under
the MIT license.

% ==================================================================
\bibliographystyle{IEEEtranN}
\bibliography{references}
% ==================================================================

\end{document}